\documentclass[english]{article}
\usepackage[T1]{fontenc}
\usepackage[latin9]{inputenc}
\usepackage{mathrsfs}
\usepackage{amsmath}
\usepackage{amssymb}
\usepackage{esint}

\makeatletter

\newcommand{\noun}[1]{\textsc{#1}}

\usepackage{babel}

\makeatother

\usepackage{babel}
\begin{document}
\title{Connecting two stochastic theories that lead to quantum mechanics}
\author{Luis de la Peña, Ana María Cetto and Andrea Valdés-Hernández \\
Instituto de Física, Universidad Nacional Autónoma de México}
\maketitle
\begin{abstract}
The connection is established between two theories that have developed
independently with the aim to describe quantum mechanics as a stochastic
process, namely stochastic quantum mechanics (\noun{sqm}) and stochastic
electrodynamics (\noun{sed}). Important commonalities and complementarities
between the two theories are identified, notwithstanding their dissimilar
origins and approaches. Further, the dynamical equation of \noun{sqm}
is completed with the radiation terms that are an integral element
in \noun{sed}. The central problem of the transition to the quantum
dynamics is addressed, pointing to the key role of diffusion in the
emergence of quantization. 
\end{abstract}

\section{Introduction}

In this paper we pay attention to two different theories that have
been successfully developed with the purpose of describing the quantum
phenomenon as a stochastic process. On one hand we have stochastic
quantum mechanics, \noun{sqm} (also known as stochastic mechanics),
a phenomenological theory initiated by Edward Nelson, and further
developed and extended independently by several groups; a representative
sample of related works is provided in Refs. \cite{Nels66}-\cite{Dice}
Ch. 2 and references therein. On the other hand we have stochastic
electrodynamics, \noun{sed}, a first-principles theory pioneered by
Trevor Marshall \cite{Mars63,Mars65} and further developed and completed
with the contributions from a number of other authors, as shown in
Refs. \cite{Dice,Clave81}-\cite{TEQ} and references contained therein.
A common feature of these theories is the explicit introduction of
stochasticity as an ontological element missing in the quantum theory,
with the aim to address many of the historical ---and still current---
conceptual difficulties associated with quantum mechanics.

In both these theories the dynamics of a representative particle of
mass $m$ is considered, for simplicity. In the phenomenological approach
of \noun{sqm} the (statistical) concepts of a flux velocity $\boldsymbol{v}$
and a diffusive velocity $\boldsymbol{u}$ are introduced on an equal
footing, without the need to specify the source of stochasticity.
A generic equation of motion is obtained, which serves to describe
the dynamics of two distinct types of stochastic process, in the Markov
approximation: the classical, Brownian-motion type and the quantum
one. The mathematics are simple and straightforward, and their physical
meaning is clear. 

The approach of \noun{sed}, on the other hand, is guided by the hypothesis
of the existence of the (random) zero-point radiation field, \noun{zpf}.
This rather more elaborate approach goes through a statistical evolution
equation (a generalized Fokker-Planck-type equation, \noun{gfpe})
in phase space, to arrive at a description in $\boldsymbol{x}$-space,
in which the dissipative and diffusive terms are seen to bring about
a definitive departure from the classical Hamiltonian dynamics. The
interplay between these two terms is what allows the system to eventually
reach equilibrium and thus attain the quantum regime; the dynamics
is then described by the Schrödinger equation, and the operators become
a natural tool for its description. Planck's constant enters into
the picture through the spectral density of the \noun{zpf}, and this
allows to determine uniquely the value of the only free parameter
introduced in \noun{sed}, as well as in \noun{sqm}.

The purpose of the present work is to establish the connection between
\noun{sqm} and \noun{sed} and, by so doing, to identify the strengths
and limitations of the two theories, as well as certain commonalities
and complementarities between them. With this aim, we first present
the basic elements of \noun{sqm} leading to the dynamical law that
governs both classical and quantum stochastic processes in the Markov
approximation. Secondly, we briefly review the statistical treatment
followed in \noun{sed} to arrive at a description in configuration
space, and discuss the conditions under which the system attains equilibrium
and thus reaches the quantum regime as described by the Schrödinger
equation, which corresponds to the radiationless approximation of
\noun{sed}. The discussion of the connections between the two theories
provides an opportunity to highlight the role played by diffusion
in quantum mechanics. The more complete dynamical description provided
by \noun{sed}, which includes the radiative terms, serves in its turn
to complete the corresponding dynamical equation of \noun{sqm}. The
distinct nature of the diffusive terms allows us to address the central
problem of the transition from the initially classical dynamics with
\noun{zpf}, to the quantum one. It is concluded that this more complete
ontology which includes the \noun{zpf} as the source of stochasticity,
leads in a natural process to the quantum description.

\section{The underlying equations of stochastic quantum mechanics\label{sqm}}

Stochastic quantum mechanics is a phenomenological theory that considers
a particle of mass $m$ undergoing a stochastic motion. It is general
enough as to accommodate a range of physical phenomena in which an
underlying stochastic process, considered in the Markov (second-order)
approximation, takes place. The stochastic nature of the dynamics
calls for a statistical treatment, which is carried out in $\boldsymbol{x}$-space.
The basic kinematic elements for the description are obtained by applying
an average over the ensemble of particles in the neighborhood of $\boldsymbol{x}$
at times close to $t$. By taking the time interval $\varDelta t$
small but different from zero, two different velocities are obtained,
namely the flux (or \emph{systematic}) velocity (see e. g. \cite{Nels66,Dice})
\begin{equation}
\boldsymbol{v}(\boldsymbol{x},t)=\frac{\overline{\boldsymbol{x}(t+\Delta t)-\boldsymbol{x}(t-\Delta t)}}{2\Delta t}=\mathcal{\hat{D}}_{c}\boldsymbol{x},\label{12.7}
\end{equation}
with 
\begin{equation}
\mathcal{\hat{D}}_{c}=\frac{\partial}{\partial t}+\boldsymbol{v}\cdot\boldsymbol{\nabla},\label{A12a}
\end{equation}
and the diffusive (or \emph{osmotic}) velocity 
\begin{equation}
\boldsymbol{u}(\boldsymbol{x},t)=\frac{\overline{\boldsymbol{x}(t+\Delta t)+\boldsymbol{x}(t-\Delta t)-2\boldsymbol{x}(t)}}{2\Delta t}=\mathcal{\hat{D}}_{s}\boldsymbol{x},\label{12.10}
\end{equation}
with 
\begin{equation}
\mathcal{\hat{D}}_{s}=\boldsymbol{u}\cdot\boldsymbol{\nabla}+D\boldsymbol{\nabla}^{2},\label{A12b}
\end{equation}
and 
\begin{equation}
D=\frac{\overline{(\Delta x)^{2}}}{2\Delta t}\label{12.12}
\end{equation}
the diffusion coefficient, assumed to be constant. The symbol $\overline{\left(\cdot\right)}$
denotes the aforementioned ensemble averaging.

By considering the forward and backward Fokker-Planck equations for
the probability density in $\boldsymbol{x}$-space $\rho(\boldsymbol{x},t)$
(see, e.g., \cite{Risk84}), and combining them appropriately, it
follows that $\rho(\boldsymbol{x},t)$ is related to the flux velocity
through the continuity equation 
\begin{equation}
\frac{\partial\rho}{\partial t}+\boldsymbol{\nabla}\cdot\left(\rho\boldsymbol{v}\right)=0,\label{12.14}
\end{equation}
and to the diffusive velocity according to 
\begin{equation}
\boldsymbol{u}(\boldsymbol{x},t)=D\frac{\boldsymbol{\nabla}\rho}{\rho}.\label{30}
\end{equation}
This most important relation confirms the diffusive meaning of the
velocity $\boldsymbol{u}$.

The two time derivatives (\ref{A12a}) and (\ref{A12b}), applied
to the velocities (\ref{12.7}) and (\ref{12.10}), give rise to four
different accelerations, thus leading to a couple of generic dynamical
equations, which are, respectively, the time-reversal invariant generalization
of Newton's Second Law, and the time-reversal non-invariant equation,
namely \begin{subequations} \label{A10} 
\begin{eqnarray}
m\left(\mathcal{\hat{D}}_{c}\boldsymbol{v}-\lambda\mathcal{\hat{D}}_{s}\boldsymbol{u}\right)=\boldsymbol{f}_{+},\label{A10a}\\
m\left(\mathcal{\hat{D}}_{c}\boldsymbol{u}+\mathcal{\hat{D}}_{s}\boldsymbol{v}\right)=\boldsymbol{f}_{-},\label{A10b}
\end{eqnarray}
where $\lambda$ is a free, real parameter, and the net force acting
on the particle $\boldsymbol{f}$ decomposes as $\boldsymbol{f}=\boldsymbol{f}_{+}+\boldsymbol{f}_{-}$,
such that $\boldsymbol{f}_{-}$ and $\boldsymbol{f}_{+}$ do and do
not change sign, respectively, under time reversal (notice that $\boldsymbol{v}$
changes its sign whereas $\boldsymbol{u}$ remains invariant).

Since equations (\ref{A10}) hold simultaneously and together they
describe the dynamics of the system, it is convenient to combine them
into a single equation. This is readily achieved by introducing the
symbol $\kappa=\sqrt{-\lambda}$ and multiplying the second equation
by $\kappa$; the result is \end{subequations} 
\begin{equation}
\mathcal{\hat{D}}_{\kappa}\boldsymbol{p}_{\kappa}=\boldsymbol{f}_{\kappa},\label{A16}
\end{equation}
with 
\begin{equation}
\boldsymbol{p}_{\kappa}=m\boldsymbol{w}_{\kappa}+\frac{e}{c}\boldsymbol{A},\quad\boldsymbol{w}_{\kappa}=\boldsymbol{v}+\kappa\boldsymbol{u},\label{A18}
\end{equation}
\begin{equation}
f_{\kappa}=\boldsymbol{f}_{+}+\kappa\boldsymbol{f}_{-},\label{A19}
\end{equation}
and 
\begin{equation}
\mathcal{\hat{D}}_{\kappa}=\mathcal{\hat{D}}_{c}+\kappa\mathcal{\hat{D}}_{s}=\frac{\partial}{\partial t}+\frac{1}{m}\boldsymbol{p}_{\kappa}\cdot\boldsymbol{\nabla}+\kappa\,D\,\boldsymbol{\nabla}^{2}.\label{A20}
\end{equation}
Equation (\ref{A16}) is the equation of motion appropriate for the
description of an ensemble of electrically charged particles immersed
in an external electromagnetic field $\boldsymbol{A}$, and subject
to stochastic forces. The Newtonian limit (or equivalently, the classical
Hamiltonian description) corresponds to $D=0$ and hence $\boldsymbol{u}=\boldsymbol{0}$,
which means no diffusion at all.

For simplicity in the derivations we shall assume no external electromagnetic
field $\boldsymbol{A}$, so that the momentum is simply $\boldsymbol{p}_{\kappa}=m\boldsymbol{w}_{\kappa}$
and the external force components reduce to 
\begin{equation}
\boldsymbol{f}_{+}=\boldsymbol{f}=-\boldsymbol{\nabla}V,\;\boldsymbol{f}_{-}=0.\label{A22}
\end{equation}

Equations (\ref{A16})-(\ref{A20}) show that the specific dynamical
properties of the system strongly depend on the sign of the parameter
$\lambda$, which in its turn determines whether $\kappa$ is real
or imaginary. Since only the sign of $\lambda$ is relevant (its magnitude
can be absorbed into the value of $D$, as explained in \cite{TEQ,CePeVaSpecif15})
one can take $\lambda=\pm1$. The value $\lambda=-1$ ($\kappa=1$)
implies an irreversible dynamics, of the Brownian-motion type. In
contrast, by setting $\lambda=1$ ($\kappa=-i$) one obtains after
some algebra the Schrödinger-like equation 
\begin{equation}
-2mD^{2}\boldsymbol{\nabla}^{2}\psi(\boldsymbol{x},t)+V(\boldsymbol{x})\psi(\boldsymbol{x},t)=2imD\frac{\partial\psi(\boldsymbol{x},t)}{\partial t},\label{A24}
\end{equation}
and its complex conjugate, where $\psi(\boldsymbol{x},t)$ is a complex
function such that 
\begin{equation}
\rho(\boldsymbol{x},t)=|\psi(\boldsymbol{x},t)|^{2}
\end{equation}
and 
\begin{equation}
\boldsymbol{v}=iD\left(\frac{\nabla\psi^{*}}{\psi^{*}}-\frac{\nabla\psi}{\psi}\right),\;\boldsymbol{u}=D\left(\frac{\nabla\psi^{*}}{\psi^{*}}+\frac{\nabla\psi}{\psi}\right),\label{A26}
\end{equation}
whence 
\begin{equation}
\boldsymbol{w}=\boldsymbol{v}-i\boldsymbol{u}=-2iD\frac{\nabla\psi}{\psi}.\label{A28}
\end{equation}

\section{The underlying equations of stochastic electrodynamics}

\subsection{The generalized Fokker-Planck Equation}

We recall that the equation of motion of \textsc{sed} for a (nonrelativistic)
particle of mass $m$ and electric charge $e$ is the Langevin equation,
also known in \noun{sed} as Braffort-Marshall equation \cite{Clave81,San81,Dice},
\begin{equation}
m\boldsymbol{\ddot{x}}=\boldsymbol{f}(\boldsymbol{x})+m\tau\boldsymbol{\dddot{x}}+e\boldsymbol{E}_{0}(t),\label{10}
\end{equation}
where $\tau=2e^{2}/3mc^{3}$, and $\boldsymbol{f}=-\boldsymbol{\nabla}V$.
The (random) electromagnetic \textsc{zpf} is usually taken in the
dipole approximation and is therefore represented by $\boldsymbol{E}_{0}(t)$.
With the momentum defined as 
\begin{equation}
\boldsymbol{p}=m\boldsymbol{\dot{x}},\label{11}
\end{equation}
Eq. (\ref{10}) transforms into 
\begin{equation}
\boldsymbol{\dot{p}}=\boldsymbol{f}+m\tau\boldsymbol{\dddot{x}}+e\boldsymbol{E}_{0}(t).\label{13}
\end{equation}
Since the dynamics of the system becomes stochastic due to the \textsc{zpf},
its evolution can only be described in statistical terms. We therefore
follow a standard procedure (see \cite{TEQ} Sec. 4.2) that leads
to the following generalized Fokker-Planck equation (\textsc{gfpe})
for the phase-space distribution $Q(\boldsymbol{x},\boldsymbol{p},t)$,
\begin{equation}
\hat{L}Q=\left(\hat{L}_{c}+e^{2}\hat{L}_{r}\right)Q=0,\label{16}
\end{equation}
where 
\begin{equation}
\hat{L}_{c}=\frac{\partial}{\partial t}+\frac{1}{m}\boldsymbol{\nabla}\cdot\boldsymbol{p}+\boldsymbol{\nabla}_{p}\cdot\boldsymbol{f}\label{17}
\end{equation}
and 
\begin{equation}
\hat{L}_{r}=\boldsymbol{\nabla}_{p}\cdot\left(\frac{m\tau}{c^{2}}\boldsymbol{\dddot{x}}-\mathcal{\hat{\boldsymbol{\mathscr{D}}}}\right).\label{18}
\end{equation}
The operator $\hat{L}_{c}$ contains the classical (i. e., conservative
and nondiffusive) Liouvillian terms, and $\hat{L}_{r}$ the radiative
and diffusive terms, the latter being represented by the integro-differential
operator $\mathcal{\hat{\boldsymbol{\mathscr{D}}}}$. To lowest order
in $e^{2}$, this operator takes the form 
\begin{equation}
\mathcal{\hat{\boldsymbol{\mathscr{D}}}}=\intop_{-\infty}^{t}dt'\varphi(t-t')\nabla_{p'},\label{22-1}
\end{equation}
where 
\begin{equation}
\varphi(t)=\frac{2\hbar}{3\pi c^{3}}\int_{0}^{\infty}d\omega\thinspace\omega^{3}\cos\omega t\label{23}
\end{equation}
denotes the \noun{zpf} covariance, and $\boldsymbol{p}'=\boldsymbol{p}(t')$
evolves towards $\boldsymbol{p}(t)$ under the action of $\hat{L}$.
Notice that it is through this diffusive term that Planck's constant
appears in the description.

\subsection{Evolution equations in configuration space\label{ee}}

From Eq. (\ref{16}) follows the equation of evolution in $\boldsymbol{x}$-space
for any dynamical variable $G(\boldsymbol{x},\boldsymbol{p})$ of
interest without explicit time-dependence, by left-multiplying the
equation by $G$ and integrating over the momentum space. The local
mean value of $G$ is 
\begin{equation}
\langle G\rangle_{x}\equiv\frac{1}{\rho}\int d\boldsymbol{p}\,G(\boldsymbol{x},\boldsymbol{p})Q(\boldsymbol{x},\boldsymbol{p},t),\label{local}
\end{equation}
where $\rho=\rho_{x}=\rho(\boldsymbol{x},t)=\int d\boldsymbol{p}\,Q(\boldsymbol{x},\boldsymbol{p},t)$
stands for the probability density. Here we consider only the results
corresponding to $G=1$ and $G=\boldsymbol{p}$. In the first case,
a direct integration of equation (\ref{16}) over $\boldsymbol{p}$
gives the continuity equation for $\rho$, 
\begin{equation}
\frac{\partial\rho}{\partial t}+\boldsymbol{\nabla}\cdot\boldsymbol{j}=0,~\boldsymbol{j}=\rho\boldsymbol{v},\label{20}
\end{equation}
with $\boldsymbol{v}=\boldsymbol{v}(\boldsymbol{x},t)$ the flux (or
current) velocity, 
\begin{equation}
\boldsymbol{v}=\frac{1}{m}\langle\boldsymbol{p}\rangle_{x}.\label{21}
\end{equation}
For $G=\boldsymbol{p}$ one gets, using (\ref{11}) and summing over
repeated indices, 
\begin{equation}
\frac{\partial}{\partial t}m\boldsymbol{v}\rho+m^{2}\partial_{j}\left\langle \dot{x}_{j}\boldsymbol{\dot{x}}\right\rangle _{x}\rho-\left\langle \boldsymbol{f}\right\rangle _{x}\rho=\boldsymbol{R},\label{22}
\end{equation}
with 
\begin{equation}
\boldsymbol{R}=m\tau\left\langle \boldsymbol{\dddot{x}}\right\rangle _{x}\rho-e^{2}\langle\mathcal{\hat{\boldsymbol{\mathscr{D}}}}\rangle_{x}\rho\label{24}
\end{equation}
containing the radiative and diffusive terms, which are of the order
of $e^{2}$.

As is shown in detail in Refs. \cite{TEQ} $\mathsection$ 4, and
\cite{PeCeVaEta14}, the left-hand side of Eq. (\ref{22}) can be
transformed into the Schrödinger-like equation 
\begin{equation}
-\frac{2\eta^{2}}{m}\boldsymbol{\nabla}^{2}\psi(\boldsymbol{x},t)+V(\boldsymbol{x})\psi(\boldsymbol{x},t)=2i\eta\frac{\partial\psi(\boldsymbol{x},t)}{\partial t}\label{Schro}
\end{equation}
with $\psi(\boldsymbol{x},t)$ a complex function such that 
\begin{equation}
\rho(\boldsymbol{x},t)=|\psi(\boldsymbol{x},t)|^{2},\label{31cc}
\end{equation}
$\eta$ a free (undetermined) parameter, and
\begin{equation}
\boldsymbol{v}(\boldsymbol{x},t)=\frac{1}{m}\text{Re}\left(\frac{-2i\eta\boldsymbol{\nabla}\psi}{\psi}\right)=-\frac{i\eta}{m}\left(\frac{\boldsymbol{\nabla}\psi}{\psi}-\frac{\boldsymbol{\nabla}\psi^{\ast}}{\psi^{\ast}}\right).\label{31c}
\end{equation}

It is important to note that neither the left-hand side of (\ref{22})
nor the resulting equation (\ref{Schro}), contain any element that
is explicitly related with the \noun{zpf} nor with radiation reaction.
In fact it is just through the balance eventually achieved between
the average energy lost by radiaton reaction and that gained from
the \noun{zpf} (the two terms deriving from the action of $\hat{L}_{r}$,
Eq. (\ref{18})), that the value of the parameter $\eta$ is determined.
It is thus found that \cite{TEQ,PeCeVaEta14}
\begin{equation}
\eta=\hbar/2,\label{eta}
\end{equation}
which transforms (\ref{Schro}) into the true Schrödinger equation;
the term on the right-hand side of Eq. (\ref{22}) represents the
radiative corrections. We shall come back to this crucial point in
Section \ref{rc}.

\section{Connecting \noun{sed} with \noun{sqm}}

\subsection{Comparing the dynamical equations}

To explore the connection between the two theories we start by noticing
that (\ref{31c}) relates the flux velocity with the real part of
the complex vector $(-i\hbar\boldsymbol{\nabla}\psi)/\psi$, while
the corresponding imaginary term, on its part, gives the velocity
vector 
\begin{equation}
\boldsymbol{u}(\boldsymbol{x},t)=-\frac{1}{m}\text{Im}\left(\frac{-i\hbar\boldsymbol{\nabla}\psi}{\psi}\right)=\frac{\hbar}{2m}\left(\frac{\boldsymbol{\nabla}\psi}{\psi}+\frac{\boldsymbol{\nabla}\psi^{\ast}}{\psi^{\ast}}\right)=\frac{\hbar}{2m}\frac{\boldsymbol{\nabla}\rho}{\rho}.\label{31d}
\end{equation}

These expressions coincide precisely with those obtained for the two
velocities of \noun{sqm}, namely Eqs. (\ref{A26}), if the diffusion
coefficient appearing in these equations is assigned the value 
\[
D=\frac{\hbar}{2m}.
\]
In \noun{sed} ---as in quantum mechanics--- $\boldsymbol{v}$ and
$\boldsymbol{u}$ represent local ensemble averages; the \noun{sqm}
expressions (\ref{12.7}) and (\ref{12.10}) represent averages over
the ensemble of particles in the neighborhood of $\boldsymbol{x}$,
which is a different way of saying the same.

In terms of these velocities, the full \textsc{sed} equation (\ref{22})
reads 
\begin{equation}
m\frac{\partial v_{i}}{\partial t}\boldsymbol{-}mv_{i}\left(\frac{2m}{\hbar}\boldsymbol{u}\cdot\boldsymbol{v}+\boldsymbol{\nabla}\cdot\boldsymbol{v}\right)+m\left(\frac{2m}{\hbar}\boldsymbol{u}\cdot\left\langle \boldsymbol{\dot{x}}\dot{x}_{i}\right\rangle _{x}+\boldsymbol{\nabla}\cdot\left\langle \boldsymbol{\dot{x}}\dot{x}_{i}\right\rangle _{x}\right)=f_{i}+\frac{1}{\rho}R_{i}.\label{32}
\end{equation}
For clarity we introduce the tensor $T_{ij}$, given by the (local)
correlation between the $i$-th and $j$-th components of the vector
$\boldsymbol{\dot{x}}$, 
\begin{equation}
T_{ij}=-\frac{2m}{\hbar}\left(\left\langle \dot{x}_{i}\dot{x}_{j}\right\rangle _{x}-v_{i}v_{j}\right)=-\frac{2m}{\hbar}\left(\left\langle \dot{x}_{i}\dot{x}_{j}\right\rangle _{x}-\left\langle \dot{x}_{i}\right\rangle _{x}\left\langle \dot{x}_{j}\right\rangle _{x}\right),\label{34}
\end{equation}
so that Eq. (\ref{32}) takes the form 
\begin{equation}
m\left(\frac{\partial v_{i}}{\partial t}-T_{ij}u_{j}-\frac{\hbar}{2m}\partial_{j}T_{ij}+v_{j}\partial_{j}v_{i}\right)=f_{i}+\frac{1}{\rho}R_{i}.\label{36}
\end{equation}
Barring the radiative corrections, represented by the last term, this
dynamical equation reduces to 
\begin{equation}
m\left(\frac{\partial v_{i}}{\partial t}-T_{ij}u_{j}-\frac{\hbar}{2m}\partial_{j}T_{ij}+v_{j}\partial_{j}v_{i}\right)=f_{i}.\label{36b}
\end{equation}

The \noun{sqm} dynamical equation (\ref{A10a}), in its turn, reads
explicitly
\begin{equation}
m\left(\frac{\partial v_{i}}{\partial t}+v_{j}\partial_{j}v_{i}-u_{j}\partial_{j}u_{i}-D\partial_{j}\partial_{j}u_{i}\right)=f_{i}\label{sqm1}
\end{equation}
in the absence of an external field, when Eq. (\ref{A22}) holds.
This coincides with the (non-radiative) dynamical equation of \textsc{sed},
Eq. (\ref{36b}), with $T_{ij}$ given by 
\begin{equation}
T_{ij}=\partial_{j}u_{i}.\label{38}
\end{equation}
Hence, by inserting (\ref{38}) into (\ref{36}) we obtain an extended
equation for \noun{sqm} that includes the radiative contributions
represented by $R_{i}$,
\begin{equation}
m\left(\frac{\partial v_{i}}{\partial t}+v_{j}\partial_{j}v_{i}-u_{j}\partial_{j}u_{i}-D\partial_{j}\partial_{j}u_{i}\right)=f_{i}+\frac{1}{\rho}R_{i}.\label{40}
\end{equation}

On the other hand, the \noun{sqm} Eq. (\ref{A10b}) leads after one
integration to the continuity equation, which is equivalent to the
\noun{sed} equation (\ref{20}). 

In this form the connection between \textsc{sqm} and \textsc{sed}
is established. The theories are seen to complement one another: while
\textsc{sqm} offers the advantage of naturally incorporating from
the beginning the couple of velocities $\boldsymbol{v}$ and $\boldsymbol{u}$
to describe the dynamics due to an (unidentified) stochastic source,
\textsc{sed} recognizes the \textsc{zpf} as the determining ingredient
that serves to precise the origin of the (quantum) fluctuations, and
introduces Planck's constant into the ultimate quantum description.
The specific value of $D$ constitutes a postulate in \textsc{sqm},
since in this theory the nature of the stochastic source remains unidentified.
Things change when making the connection of \textsc{sqm} with \textsc{sed},
since in the latter theory the \textsc{zpf} with energy per mode $\hbar\omega/2$,\textsc{\ }is
the natural carrier of $\hbar$.

\subsection{Evidence of diffusion in quantum mechanics}

A significant hint of the direct connection of \textsc{sed} and \textsc{sqm}
with quantum mechanics follows by observing that the quantum momentum
operator is directly related with the velocity $\boldsymbol{w}_{\kappa}$
for $\kappa=-i$, Eq. (\ref{A28}), 
\begin{equation}
\boldsymbol{\hat{p}}\psi=-i\hbar\boldsymbol{\nabla}\psi=(\boldsymbol{v}-i\boldsymbol{u})\psi.\label{31A}
\end{equation}
This result reveals that both velocities $\boldsymbol{v}$ and $\boldsymbol{u}$
are a natural part of quantum mechanics, even if $\boldsymbol{v}$
is rarely used (see however \cite{Bal98}, and $\boldsymbol{u}$ remains
virtually ignored. In terms of these velocities, the (quantum) expectation
value of the squared momentum reads 
\begin{equation}
\left\langle \boldsymbol{\hat{p}}^{2}\right\rangle =m^{2}\left\langle \boldsymbol{v}^{2}+\boldsymbol{u}^{2}\right\rangle ,\label{31G}
\end{equation}
and the quantum variance 
\begin{equation}
\sigma_{\boldsymbol{\hat{p}}}^{2}=\left\langle \boldsymbol{\hat{p}}^{2}\right\rangle -\left\langle \boldsymbol{\hat{p}}\right\rangle ^{2}\label{31F}
\end{equation}
is given by 
\begin{equation}
\sigma_{\boldsymbol{\hat{p}}}^{2}=\sigma_{m\boldsymbol{v}}^{2}+\sigma_{m\boldsymbol{u}}^{2},\label{31H}
\end{equation}
where the variance of a generic vector $\boldsymbol{b}(\boldsymbol{x},t)$
is given by $\sigma_{\boldsymbol{b}}^{2}=\langle\boldsymbol{b}^{2}\rangle-\langle\boldsymbol{b}\rangle^{2}$,
with $\langle\cdot\rangle=\int d\boldsymbol{x}\,(\cdot)\,\rho(\boldsymbol{x},t)$. 

Since 
\begin{equation}
\sigma_{\boldsymbol{u}}^{2}=\left\langle \boldsymbol{u}^{2}\right\rangle =\int d\boldsymbol{x}\,\rho(\boldsymbol{x},t)\,\boldsymbol{u}^{2}(\boldsymbol{x},t)>0,\label{31I}
\end{equation}
momentum dispersion is unavoidable in quantum mechanics --- the single
exception being the free particle in a $\boldsymbol{p}$-eigenstate,
in which case the position dispersion is infinite. A well-known manifestation
of this is the Heisenberg inequality $\Delta x\Delta p\geq\hbar/2$. 

Another distinctive and persisting manifestation of the diffusive
velocity $\boldsymbol{u}$ is the so-called quantum potential,
\begin{equation}
V_{Q}=-\hbar^{2}\left(\nabla^{2}\sqrt{\rho}\right)/2\sqrt{\rho}=-\frac{1}{2}\left(m\boldsymbol{u}^{2}+\hbar\boldsymbol{\nabla}\cdot\boldsymbol{u}\right).\label{45}
\end{equation}
This energy contribution totally due to fluctuations is of paramount
importance in determining much of the quantum behavior; we recall
that it plays a central role in Bohm's interpretation of quantum mechanics
\cite{BoHiUndivided93}.

Along the present discussion we have met the confluence of both theories,
\noun{sqm} and \noun{sed, }with quantum mechanics, through the equivalence
of their statistical nature as being described by the Schrödinger
equation. But there is more, since results such as (\ref{31A})-(\ref{31H})
furnish convincing evidence that, along with the Schrödinger equation,
the whole Hilbert-space formalism is involved in such correspondence.

\section{The mechanism of the classical-to-quantum transition \label{c2q}}

\subsection{Radiation and diffusion}

Let us now pay attention to the radiative contributions, represented
by the term $e^{2}\hat{L}_{r}Q$ in the \textsc{gfpe} (\ref{16}).
For this purpose we multiply this equation by any constant of motion
$G(\boldsymbol{\boldsymbol{x}},\boldsymbol{\boldsymbol{p}})=\xi$
and integrate over \emph{$\boldsymbol{p}$}. The terms associated
with the classical Liouvillian, $\hat{L}_{c}\xi$, cancel out automatically,
and only the two terms associated with $\hat{L}_{r}\xi$ remain. For
equilibrium to be reached, these terms must eventually balance each
other. By resorting to Eqs. (\ref{22-1}) and (\ref{23}), one obtains
for the balance condition 
\begin{equation}
-\left\langle \dddot{\boldsymbol{x}}\cdot\boldsymbol{g}\right\rangle _{x}=\frac{\hbar}{2}\int_{0}^{\infty}d\omega\,\omega^{3}\int_{-\infty}^{t}dt\cos\omega(t-t^{\prime})\left\langle \nabla_{p^{\prime}}\cdot\boldsymbol{g}\right\rangle _{x},\label{10-1}
\end{equation}
with $\boldsymbol{g}(\boldsymbol{x},\boldsymbol{p})=\nabla_{p}\xi(\boldsymbol{x},\boldsymbol{p})$
and $\boldsymbol{p}^{\prime}=\boldsymbol{p}(t^{\prime}),$ $t^{\prime}<t$.

Although the equality in (\ref{10-1}) holds only under equilibrium,
each side of it can be analyzed separately for all times. It is clear
that the two terms reflect different dynamical properties of the system.
Whereas initially (at $t=-\infty$, when particle and \textsc{zpf}
start to interact and there is no diffusion) the radiation term (left-hand
side) obviously dominates over the diffusive one (right-hand side),
with time the diffusion of the momentum increases due to the action
of the \textsc{zpf}. Thus, while the system starts from a nonequilibrium
condition, the two dynamical processes allow it to converge towards
a balance regime in which the $\xi$ are indeed constant.

Fundamental to the analysis is the factor $\nabla_{p^{\prime}}\cdot\boldsymbol{g}$,
which is at the core of the mechanism of evolution towards the balance
regime. This coefficient signals the effects on $\boldsymbol{g}(\boldsymbol{x},\boldsymbol{p})$,
of the diffusion of the particles activated by the \textsc{zpf} through
its direct action on the momentum $\boldsymbol{p}$. In classical
mechanics, the quantity $\nabla_{p^{\prime}}\cdot\boldsymbol{g}$
can be expressed in terms of a Poisson bracket involving $\boldsymbol{g}$
at time $t$ and $\boldsymbol{p}$ at time $t^{\prime}$, 
\begin{equation}
\frac{\partial g_{i}}{\partial p_{j}'}=\left[x_{j}',g_{i}\right].\label{50-1}
\end{equation}
The Poisson bracket represents an abridged description of the Hamiltonian
evolution, controlled by the classical Liouvillian $L_{c}$; in this
case the dynamics is purely deterministic. By contrast, the dynamics
contained in equation (\ref{10-1}) is controlled by the entire Liouvillian,
and is therefore deterministic in a statistical sense only. This means
that although the motion of each particle follows deterministic rules,
the fact that it is acted upon by a stochastic field makes the evolution
of the ensemble of particles \emph{statistically deterministic}, hence
not amenable to a purely Hamiltonian description. The conforming (modified)
Newton equations of motion are of a nature akin to that discussed
in section \ref{sqm} and reflected in Eq. (\ref{A16}), which appropriately
incorporates the effects of diffusion. As a consequence, the right-hand
side of Eq. (\ref{10-1}) ---and with it the entire equation---
ceases to obey Hamiltonian laws as soon as the diffusion enters into
force. 

In conclusion, although initially the dynamics is controlled by Hamiltonian
laws, as the interaction develops diffusion eventually takes control.
At this point Eq. (\ref{10-1}) acquires validity, signaling the passage
to classical+\noun{zpf} physics in the balance regime. The new laws,
which are statistical in nature by virtue of the action of the \noun{zpf,}
coincide with those of quantum mechanics. This means that the Poisson
brackets have been replaced by their corresponding commutator. The
presence of $\hbar$ in the commutator provides an important clue
---although rarely appreciated if at all in its daily use: it is
a direct result of the crucial role played by the \noun{zpf} in the
dynamics, and evinces the transition from initially conventional classical
to classical+\noun{zpf} physics, and eventually to \noun{sed} in the
balance regime, i. e. to quantum physics. That this qualitative change
stems from an underlying physical mechanism of transition mastered
by the \noun{zpf}, may sound natural to some, radical to others; in
fact, it is both. Interestingly, however, a qualitative change due
to a transition from an initially classical dynamics into one which
is fundamentally quantum in nature, has already been observed in experiments
with open photonic systems \cite{Raft14}.

\subsection{Two brands of stochastic processes}

In the general approach to \textsc{sqm} as briefly discussed in section
\ref{sqm} (and more extensively in, e. g., \cite{Pe69,TEQ,CePeVaSpecif15}),
the description of the dynamics involves the undetermined coefficient
$\lambda$ that can take the values $+1$ or $-1$, thereby opening
the way to the study of two essentially different dynamics. Indeed,
this parameter defines the sign of an acceleration related to the
diffusion that is to be either added or subtracted to the drift-related
acceleration (as shown in Eq. (\ref{A10a})), so that the dynamical
laws differ from one another, and from the classical (Newtonian) law,
due precisely to the diffusive terms. In the referred works and as
discussed above, it is shown that the selection $\lambda=-1$ corresponds
to Brownian motion, whereas $\lambda=1$ leads to quantum mechanics
(through the Schrödinger equation). The close relationship between
\textsc{sqm} and \textsc{sed} shows that, despite their dissimilarities,
both stochastic processes share certain laws, such as equations (\ref{A16}-\ref{A22}).

A natural question that emerges from the previous discussion is, how
is it that the transition to quantum mechanics occurs in the \textsc{sed}
system but not in the case of Brownian motion, which is the most characteristic
classical stochastic process? There are several physical features
that distinguish the two stochastic processes, a first obvious one
being the scale. Whereas Brownian systems are normally microscopic
or macroscopic in size, the quantum ones are of atomic or subatomic
size, and many orders of magnitude more sensitive to the relatively
high intensity of the stochastic background ---in this case the \textsc{zpf}---,
which induces significant fluctuations on the dynamical variables
of the system. This difference in the response is so noticeable that
one of the first quantum rules to be established (already during 1927)
were the Heisenberg uncertainty relations, which in the present understanding
express properties of \emph{causal} stochastic motions, rather than
the familiar ``inherent''\ indeterminism. But of course the most
important difference refers to the source of the stochasticity, which
in the Brownian system is a white noise, free of any self-correlation,
whereas in the quantum case it is, according to our description, an
intense colored field (due to its $\omega^{3}$-spectrum) with important
spatial and temporal self-correlations. In fact, as has been shown
in the relevant literature (see \cite{TEQ} and references therein),
it is the radiation field endowed with these high correlations that
can be identified as the source of the (statistical) wavelike behavior
of quantum particles.

\subsection{Precising the ontology of quantum mechanics}

The question of whether the dynamics of a system can transit from
classical to quantum may result misleading or baffling if taken loosely.
A legitimate answer requires that the starting theory contain already
the ontological elements proper of quantum mechanics. Now, the miscellany
of conceptual problems and difficulties that beset conventional quantum
mechanics, when closely looked at, point towards the possibility of
a common origin, namely some critical component that has been left
aside. Here we are proposing to consider the zero-point radiation
field as the key missing element in the quantum ontology, and the
transition, therefore, not from plain classical physics but from classical-plus-zero-point-radiation-field
physics to quantum physics. As seen from the above analysis, this
more complete ontology leads in a natural process to the quantum description.

Equation (\ref{Schro}), along with Eqs. (\ref{31A})-(\ref{31H}),
imply that once the balance (or quantum) regime is established, the
dynamical variables can legitimately be treated via the corresponding
usual operators in Hilbert space. This perspective stands in contrast
with the historical one, in which the founders of the theory felt
compelled to introduce operators (in their matrix representation)
to account for the observed facts ---just as Newton's law of gravitation
was proposed to save the phenomenon--- without any acknowledgment
(nor knowledge) of the role played by the underlying cause, the theoretical
weight of which remained ---and still remains--- largely unrecognized,
adding its part to the opacity of quantum mechanics.

The perspective to be drawn from these results is that the \textsc{zpf}
not only plays a significant role in explaining quantum indeterminism
as the result of an induced stochasticity, but that its presence provides
the basis for an explanation of the quantum behavior of matter altogether.
(A more extensive discussion and substantiation of these matters is
presented in \cite{TEQ}.)

A note about the reverse transition from quantum to classical seems
appropriate at this point. It is usual to consider classical physics
as a limiting case of the quantum description, attained e. g. by allowing
$\hbar$ to go to zero, with the argument that in this limit all operators
commute. This is however a formal transition; the fundamental difference
in the nature of the classical-vs-quantum dynamics demands a more
in-depth consideration of this apparently simple 'change of scale'.

\subsection{Some words about the radiative corrections\label{rc}}

Equation (\ref{40}) is the dynamical law of both \noun{sed} and \textsc{sqm,
}including the radiative corrections to second order in $e$. This
more complete description allows one to obtain several important results
pertaining to the realm of quantum electrodynamics (in the non-relativistic
approximation), such as the formulas for the Einstein coefficients,
which determine in particular the lifetimes of atomic states. The
corresponding calculations and results can be seen in \cite{TEQ}
and references therein. In the context of the present work, the most
interesting application has been the determination of the diffusion
coefficient $D$ of \textsc{sqm}.

We recall that according to the \textsc{sed} equation (\ref{10-1}),
a balance must be achieved in the quantum regime between the radiative
and dissipative effects on the dynamics. In particular, for $\xi=\boldsymbol{p}^{2}/2m+V$,
Eq. (\ref{10-1}) represents the energy-balance condition, meaning
that the mean power absorbed by the particle from the \textsc{zpf}
is compensated by the mean power radiated by the former. While the
radiation reaction term contains parameters deriving from (classical)
electrodynamics only, Planck's constant enters into the second term
through the spectral energy density of the \textsc{zpf}. A detailed
calculation of the two terms shows that it is precisely this balance
condition what fixes uniquely the value of the free parameter $\eta$
used in Section \ref{ee}, and hence of the diffusion coefficient
$D$ in terms of $\hbar$.

\section{Final remarks and considerations}

For several decades already, two theories have coexisted which arrive
at quantum mechanics from an (assumed) classical context that includes
stochasticity as an essential ingredient. Historically they were developed
by different and virtually independent clusters of researchers, with
little intersection. Hence their coexistence has been more than peaceful.
Also their philosophies are quite distant, \textsc{sqm} having been
conceived of as a Brownian-type theory for the particle subject to
a \emph{white noise} from an unidentified source. By contrast, \textsc{sed}
has been developed as a statistical description for the particle subject
to the \noun{zpf} with a \emph{colored} spectrum. As shown here, the
two theories complement each other and both lead to the Schrödinger
equation after appropriate workings; thus, in the global scenario
quantum mechanics emerges from a classical$+$stochastic context.
Leaving aside the theoretical body here developed, one could ask,
why so?

The reason for the success of such parallel constructs is traced to
the role played by diffusion. In \textsc{sqm} the velocity $\boldsymbol{u}$
is introduced from the very start as a dynamical variable that encapsulates
the diffusive effect of the random force on the particle motion. Both
the diffusive velocity $\boldsymbol{u}$ and the flux velocity $\boldsymbol{v}$
are of course statistical concepts, and together with the ensuing
four accelerations they modify Newton's Second Law in an essential
way. Also \textsc{sed} starts by considering the appropriate statistical
description by means of the \textsc{gfpe}, which ensues from the (stochastic)
Langevin-type equation ---equally modifying the Second Law in an
essential way.

In this work we have established the equivalence between the equations
of motion derived in \textsc{sed} ---a fundamental theory--- and
those of \textsc{sqm} ---a phenomenological theory. One may say that
\textsc{sqm} becomes thus \emph{explained }by \textsc{sed}, and completed
by it. This is reinforced by recalling that the value of the diffusion
constant $D=\hbar/2m$ ---a free postulate in \textsc{sqm}, which
has no natural place for Planck's constant--- is derived from a consideration
of the radiative terms of \textsc{sed}, as explained in section (\ref{rc}).

A final important point is that the coherence between the \textsc{sqm}
and \textsc{sed} theories discloses in the former the presence of
an undulatory element, which it lacks of in its usual strictly corpuscular
treatments (by Nelson and followers; see \cite{Nels66,Guerra81}).
This provides a natural answer to the well-known objection against
\textsc{sqm} by Wallstrom \cite{Wall89,Wall94}, who deems the known
derivations incorrect, based on the argument that they require an
ad hoc (wave-like) quantization condition on the velocity potential
(the gradient of which gives the velocity $\boldsymbol{v})$ in order
to derive Schrödinger's equation; of course such condition appears
artificial in a strictly corpuscular framework, but it acquires a
natural place in a theory that embodies a radiation field as its substantial
source of stochasticity.


\begin{thebibliography}{10}
\bibitem{Nels66}E. Nelson (1967) \emph{Dynamical theories of Brownian
motion}, Princeton University Press, Princeton, NJ.

\bibitem{Pe69}L. de la Peña (1969) New formulation of stochastic
theory and quantum mechanics, \emph{J. Math. Phys}. \textbf{10}, 1620.

\bibitem{Guerra81}F. Guerra (1981) Structural aspects of stochastic
mechanics and stochastic field theory, \emph{Phys. Rep}. \textbf{77},
263.

\bibitem{Nels85b}E. Nelson (1985) \emph{Quantum fluctuations}, Princeton
University Press, Princeton, NJ.

\bibitem{Nels12}E. Nelson (2012) Review of stochastic mechanics,
\emph{JPCS} \textbf{361}, 012011.

\bibitem{Dice}L. de la Peña and A. M. Cetto (1996) \emph{The Quantum
Dice}, Kluwer Academic Publishers, Dordrecht, NL

\bibitem{Mars63}T. W. Marshall (1963) Random electrodynamics, \emph{Proc.
Roy. Soc}. A \textbf{276}, 475.

\bibitem{Mars65}T. W. Marshall (1965) Statistical electrodynamics,
\emph{Proc. Camb. Phil.} Soc. \textbf{61}, 537.

\bibitem{Boyer}T. H. Boyer (1975) Random electrodynamics: The theory
of classical electrodynamics with classical electromagnetic zero-point
radiation, \emph{Phys. Rev.} D \textbf{11}, 790.

\bibitem{Clave81}P. Claverie (1981) Stochastic electrodynamics versus
quantum theory: recent advances in the study of non-linear systems,
\emph{Proc. Einstein Centennial Symposium on Fundamental Physics},
S. M. Moore et al, eds. Universidad de los Andes, Bogotá, Col., p.
229.

\bibitem{San81}E. Santos (1981) Locality in stochastic electrodynamics,\emph{
Proc. Einstein Centennial Symposium on Fundamental Physics}, S. M.
Moore et al, eds. Universidad de los Andes, Bogotá, Col., p. 213.

\bibitem{PeCe77}L. de la Peña and A. M. Cetto (1977) Derivation of
quantum mechanics from stochastic electrodynamics, \emph{J. Math.
Phys.} \textbf{18}, 1612.

\bibitem{TEQ}L. de la Peña, A. M. Cetto and A. Valdés-Hernández (2015)
\emph{The Emerging Quantum}, Springer Verlag, Berlin.

\bibitem{CePeVaSpecif15}A. M. Cetto, L. de la Peña and A. Valdés-Hernández
(2015) Specificity of the Schrödinger equation, \emph{Quantum Stud.:
Math. Found}. \textbf{2}, 275.

\bibitem{Risk84}H. Risken (1984) \emph{The Fokker-Planck equation.
Methods of solution and applications}, Springer Verlag, Berlin.

\bibitem{PeCeVaEta14}L. de la Peña, A. M. Cetto and A. Valdés-Hernández
(2014) The zero-point field and the emergence of the quantum, \emph{Int.
J. Mod. Phys}. E \textbf{23:9}, 1450049.

\bibitem{Bal98}L. E. Ballentine (1998) \emph{Quantum Mechanics: A
Modern Development}, World Scientific, Singapore. 

\bibitem{BoHiUndivided93}D. Bohm and B. J. Hiley (1995) \emph{The
Undivided Universe}, Routledge, New York, NY.

\bibitem{Raft14}J. Raftery, D. Sadri, S. Schmidt, H. E. Türeci and
A. A. Houck (2014) Observation of a dissipation-induced quantum transition,
\emph{Phys. Rev.} X \textbf{4}, 031043.

\bibitem{Wall89}T. C. Wallstrom (1989) On the derivation of the Schrödinger
equation from stochastic mechanics, \emph{Found. Phys. Lett}. \textbf{2},
113.

\bibitem{Wall94}T. C. Wallstrom (1994) Inequivalence between the
Schrödinger equation and the Madelung hydrodynamic equation, \emph{Phys.
Rev.} A \textbf{49}, 1613.
\end{thebibliography}
\end{document}